\documentclass{svproc}

\usepackage{url}
\usepackage{graphicx}
\usepackage{eurosym}
\usepackage[mathscr]{eucal}
\usepackage{float}
\usepackage{color}
\usepackage{hyperref}
\usepackage[utf8]{inputenc}
\usepackage[english]{babel}
\usepackage{amsmath, amssymb, latexsym, amscd, amsfonts,amstext}

\newcommand\blfootnote[1]{%
  \begingroup
  \renewcommand\thefootnote{}\footnote{#1}%
  \addtocounter{footnote}{-1}%
  \endgroup
}

\begin{document}
\mainmatter              %
\title{Application of Convolutional Neural Networks with Quasi-Reversibility Method Results for Option Forecasting}
\titlerunning{CNN for Option Forecasting}  
\author{Zheng Cao\inst{1}, Wenyu Du\inst{2}, \and
Kirill V. Golubnichiy\inst{3}}

\authorrunning{Cao, Du, Golubnichiy} 

\institute{University of Washington, Seattle, WA 98195, USA\\
\email{zc68@uw.edu},\\
\and
University of Washington, Seattle, WA 98195, USA\\
\email{wenyudu@uw.edu},\\ WWW home page:
\texttt{https://www.duduncan.com}
\and
University of Washington, Seattle, WA 98195, USA\\
\email{kgolubni@uw.edu}
}

\maketitle              %
\blfootnote{Cao and Du contributed equally to this work.}

\begin{abstract}

This paper presents a novel way to apply mathematical finance and machine learning (ML) to forecast stock options prices. Following the previous results, we create and evaluate new empirical mathematical models for the Black-Scholes equation to analyze data for 92,846 companies. We solve the Black-Scholes (BS) equation forwards in time as an ill-posed inverse problem, using the Quasi-Reversibility Method (QRM), to predict option price for the future one day. For each company, we have 13 elements including stock and options’ daily prices, volatility, minimizer, etc. Because the market is so complicated that there exists no perfect model, we apply ML to train algorithms to make the best prediction. The current stage of research combines QRM with Convolutional Neural Networks (CNN), which learn information across a large number of data points simultaneously. We implement CNN to generate new results by validating and testing on sample market data. We test different ways of applying CNN and compare our CNN models with previous models to see if achieving a higher profit rate is possible. 

$$
$$

\end{abstract}

\section{Introduction}
For the past few decades, scholars from mathematics, economics, and computer science fields have been devoted to finding different approaches to predict the economic market.
The economic market data is known for its chaotic pattern and randomness, and with no rigorous math theories furnished to accurately formulate the growth trend prediction, we decide to use machine learning algorithms to train models to study the historical data and make predictions. Several stock option prediction models have been created by Dr. Golubnichiy and his colleagues, Dr. Mikhail V. Klibanov and Dr. Andrey V. Nikitin: Quasi-Reversibility Method (QRM), Binary Classification, and Regression Neural Network ML (Regression NN)\cite{SH}, in the paper \textit{Quasi-Reversibility Method and Neural Network Machine Learning to Solution of Black-Scholes Equations} (appeared on the \textit{AMS Contemporary Mathematics} journal). QRM is an analytical and analytical approach to finding the minimizer by solving the Black-Scholes equation as an ill-posed inverse problem; whereas the latter two combine QRM with ML to improve the result precision. 

In this paper, on the other hand, we apply a machine learning method, Convolutional Neural Network (CNN), to help improve stock prediction performances. Begin with laying out the QRM mechanics and previous models’ results, the paper introduces a detailed road map of the modeling of the CNN program from the choice of features from stock option data to data set distribution, to the building of the inner architecture of this, neural network, and gradually to the analysis and evaluation of different approaches, constraints, and future developments and applications. 
Suppose our CNN model successfully forecasts prices for a majority of stock options, it is possible to deploy the model in the real world and help investors make better investment decisions.

\subsection{Notice:}
This research is for academic study only, not for any financial applications or advice.

\section{Previous Work}
Klibanov, Kuzhuget, and Golubnichiy\cite{KlibGol} developed a new empirical mathematical way of modeling and producing more accurate stock option prices with initial and boundary conditions. The Black-Scholes (BS) equation is used when we are given time in the past to determine current value\cite{Hull}. For financial mathematics, the BS equation is a parabolic partial differential equation that targets the European style options\cite{Shreve}. Because we are trying to predict future option prices in this research, this falls into ill-posed problems. Ill-posed problem is where the solution does not exist, or the solution is unstable.

The solution to the ill-posed problem is Regularization, and the key is to convert the system into a linear form of functional.

$u$ is the minimizer, thus it is our prediction. We have a vector $X$, which contains 13 elements including the previous minimizer u and the volatility coefficient ${\sigma}$ \cite[Chapter 7, Theorem 7.7]{Bjork}\textbf{:}

\begin{equation}
\begin{split}
& \frac{\partial u(s,\tau )}{\partial \tau }=\frac{{\sigma }^{2}}{2}s^{2}\frac{%
\partial ^{2}u(s,\tau )}{\partial s^{2}}, \\
& u(s,0)=f(s),
\end{split}
\label{1}
\end{equation}%
The payoff function is $f(s)=\max \left(
s-K,0\right) $, at T = t, where $K$ is the strike price \cite{Bjork}, $s>0.$, and the time at a given time $t$ will occur is $\tau$ 
\begin{equation}
\tau =T-t. 
\end{equation}
The option price function is defined by the Black-Scholes formula: 
\begin{equation}
u(s.\tau )=s\Phi (\theta _{+}(s,\tau ))-e^{-r\tau }K\Phi (\theta _{-}(s,\tau
)),
\end{equation}%

Based on the It\^o formula, we have:
\begin{equation}
du = (-\frac{\partial u(s, T-t)}{\partial \tau} + \frac{{\sigma}^2}{2} s^2 \frac{%
\partial^2 u(s, T-t)}{\partial s^2})dt + \sigma s \frac{\partial u(s, T-t)}{%
\partial s} dW.
\end{equation}
If equation (3) is solved forwards in time to forecast prices of stock options, it is an
ill-posed inverse problem. By solving the previous equation, we get the solution as the minimizer and apply it to the trading strategy to generate prediction results. After testing on real market data, we prove all these new mathematical approaches suggest better prediction results which may help traders make better investment decisions.

Table 1 summarizes the results of QRM, Binary Classification, and Regression Neural Network Machine Learning. While the Precision column of the table indicated the percentage of profitable options under the model generations.

\bigskip \vspace{0.75cm} 
\textbf{Table 1. Previous Models' Results}
\begin{center}
\begin{tabular}{|l|l|l|l|}
\hline
Method & Accuracy & Precision & Recall \\ \hline
QRM & 49.77\% & 55.77\% & 52.43\% \\ \hline
Binary Classification & 56.36\% & 59.56\% & 70.22\% \\ \hline
Regression NN & 55.42\% & 60.32\% & 61.29\% \\ \hline
\end{tabular}
\end{center}

From these results, it is evident that QRM is promising. However, when combined with QRM, the two machine learning models, Binary Classification and Regression NN, both improved on the results of QRM. Therefore, we decided to explore a different neural network, convolutional neural networks, with results from QRM. Whereas Binary Classification and Regression NN make predictions for each stock option independently, CNNs are better suited to take into consideration multiple stock options at a time when making predictions. This will allow information about the larger market to be included in the prediction.

Our task is to implement a new machine learning algorithm with the result of the QRM using CNN.

\section{CNN Modeling}

Because the stock option market is affected by nearly every aspect of society and it is impossible to make a perfect prediction model for the future economic market, we apply machine learning (ML) to train algorithms to help improve the prediction. ML can automatically improve through experience and data. 

We input 13 elements for each company as parameters, which include the stock, ask, and bid prices for today plus the option, ask, and bid price and volatility for today and the past 2 days, and the minimizer obtained from the QRM for today and tomorrow. Therefore, for the total 92,846 companies, we have the training, validating, and testing data vectors, each representing a selected company for the given set time frame. The input data are split into 3 categories: 75.74\% training, 10.64\% validating, and 13.62\% testing, for a total of $92,846 \times 13 = 1,206,998$ values.
 
\subsection{Sample Data Set Images}

As the algorithm features we train the ML model with, the sample data we utilized includes 31 columns: option name, grid count, beta, date, option ask +2, option ask0, option bid -2, option bid -1, option bid -1, option bid0, stock ask0, stock bld0, ivol -2, ivol +1, ivol -0, est +1, est +2, option mean +1, option mean +2, option ask +1, option ask +2, option bid +2, stock ask -1, stock bid -2, stock bid +1, stock bid +1, ivol +1, ivol +2, option type, minimizer error +1, and minimizer error +2. The "+" and "-" here refer to the relative days of the present day, e.g. "stock ask -2" means the stock ask price of the day before yesterday.

Note that we have 32 columns for the data frame, but only 13 elements (features) for each option on the given day will be considered as the input. All these data are real-world stock option information between late 2019 and early 2021 imported Bloomberg.com\cite{Bloomberg}.

\subsection{Table 2. CNN Data Set Distribution}

\begin{center}
\begin{tabular}{ |p{4cm}||p{4cm}|}
\hline
Data Set Types & Total Number \\
\hline
Training Data & (70,322, 13) \\
\hline
Validation Data & (9875, 13)  \\
\hline
Testing Data & (12649, 13) \\
\hline
\end{tabular}
\end{center}
\bigskip

Table 2 is the CNN Data Set Distribution summarizing the input vectors we train, validate, and test for the machine learning algorithm. The first value is the number of data points, the second is the number of features.

We first train the algorithms using Gradient Descent, then tune the hyperparameters during the validation stage, and lastly evaluates the model precision and accuracy with the test samples. Once we finish training, we test, one and only once. Changing the model after testing is prohibited because if we modify the data based on the test result, the model will not be accurate. One example of validating hyperparameters is the prediction threshold. If the model output is greater than or equal to this threshold, we predict that this stock goes up. We create different models based on the choice of hyperparameters, and the validation data set is used to help evaluate and select the best parameter combination.  

Currently, we have two approaches for building up the Convolutional Neural Network (CNN) as a substitution for Regression Neural Network to help predict stock options. CNN Approach 1 is finished and the below section walks through the modeling procedure; while the second approach is introduced for future developments.

\subsection{Machine Learning Input Vector Normalization}

Before diving into detailed modeling procedures, we first normalize the input vector as the following equations:

\begin{equation}
\mu = \frac{u_{a}(0)+u_{a}(-\tau)+u_{a}(-2\tau)+u_{b}(0)+u_{b}(-%
\tau)+u_{b}(-2\tau)}{6}
\end{equation}

\begin{equation}
op_{n}= \frac{u(t)-\mu}{\sigma}
\end{equation}

\begin{equation}
s_{n}= \frac{(s-s_t)-\mu}{\sigma}
\end{equation}
where $op_{n}$ is a normalized option price, $s_{n}$ is a normalized stock
price normalization, $s$ is the stock, $s_t$ is the strike and $\sigma$ is
the standard deviation.

\subsection{CNN Approach 1.0:}

We choose n companies each with 13 pieces of information (including the normalized features above) as the input data, and include 6 convolution layers plus 1 max pool layer, and produce an output of n by 1 vector of (0, 1). Which we simultaneously study all companies from the input data set, and the algorithm seeks to establish a hidden relationship, thus making a corresponding trading strategy. We input a total of 70322 companies as the training data. 

\subsection{CNN Approach 1.1:}
Upon realizing that the result of CNN Approach 1 was extremely one-sided, we modified the last part of the machine learning function and use Sigmoid function. After around 100 epochs, training loss became stable. Figure 1 shows the evolution of training loss over epochs.

The final CNN model includes 5 convolutional layers, each with a 2-by-1 reflexive padding to maintain dimensionality. The result of the fifth layer is passed through a max pooling layer before normalizing with the sigmoid function. 

\begin{center}
\hspace{20pt}
\includegraphics[scale=0.7]{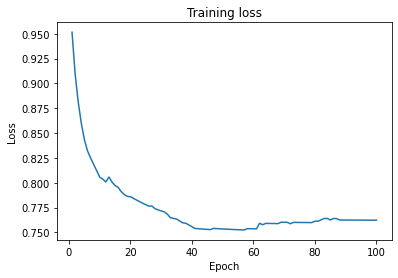}\\
Figure 1: Gradient Descent Training Loss
\end{center}

\vspace{-10pt}
\hspace{10pt}
The threshold $c$ is a hyperparameter that we tune to optimize the model. Any output smaller than $c$ will be classified as 0; all other outputs between $c$ and 1 will be classified as 1. The threshold itself has no direct correlation with the training loss above. Using the validation data set, we obtain 0.54 as the threshold value that produces the maximum accuracy.

The confusion matrix below visualizes the true labels vs. predicted labels when evaluating the test data set. From the confusion matrix, we can see that this model predicts more negatives than it does positives. However, when the model predicts a positive, it appears more accurate, indicated by the percentage of positive predictions which are correct.

\begin{center}
\hspace{20pt}
\includegraphics[scale=0.7]{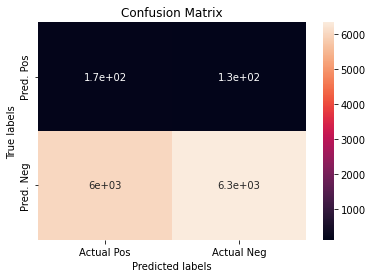}\\
Figure 2: Confusion Matrix Visualization
\end{center}

When evaluating the training data with the best threshold c = 0.54, the model generated an accuracy of 51.62\%, a precision of 53.76\%, and a recall of 47.14\%, and when evaluating on testing data with the best threshold, an accuracy of 51.49\%, the precision of 57.14\%, and the recall of 2.78\% are produced. This high precision lines up with what the confusion matrix suggests. Table 3 is how the model compares to QRM and Approach 1:

\subsection{Table 3. CNN Approach 1.0 and 1.1 results}

\bigskip
\begin{center}
\begin{tabular}{ |p{5cm}|p{1.75cm}|p{1.75cm}|p{1.75cm}|p{1.75cm}|}
\hline
Method & Accuracy & Precision & Recall\\
\hline
QRM  & 49.77\% & 55.77\% & 52.43\% \\
\hline
CNN App 1.0 Testing & 51.15\%  & N/A & 0.0\%\\
\hline
CNN App 1.1 Training & 51.62\% & 53.76\% & 47.14\% \\
\hline
CNN App 1.1 Testing & 51.49\%  & 57.14\% & 2.78\%  \\
\hline
\end{tabular}
\end{center}
\bigskip

Compared to the result of the original QRM, which has an accuracy of 49.77\%, the CNN Approach significantly increases the precision, which is equivalent to the profitable option rate, to 57.14\%, and increases the accuracy by approximately 1.72\%. With such improvement, when traders use our model to make an investment to a bountiful scale and range of options, the expected percentage of profitable rate by 1.37\%, per trading unit. The outcome that the false negative is relatively high whereas the false positive is relatively low (see Figure 2: Confusion Matrix Visualization) may explain the low recall value of 2.78\%.

\subsection{CNN Approach 2:}

Recognizing both the potential in CNN and the limits of Approach 1, we propose another CNN model to explore. Differing from the first approach, we select clusters of companies, e.g. 11 companies as a set to train the algorithm. The clusters are obtained from “neighboring” companies, and each time we only make one prediction for the center company of the cluster. 

Challenges with approach 2 include choosing neighboring companies in the cluster and assigning weights to different companies of each cluster (how to indicate certain companies as more important in the prediction).

\section{Trading Strategy}
Let’s denote s as the stock price, $t$ as the time, and $\sigma(t)$ as the volatility of the option. The historical implied volatility listed on the market data of [2] is used in our particular case. We assume that $\sigma = \sigma(t)$ to avoid other historical data for the volatility. Let’s call $ub(t)$ and $ua(t)$ the bid and ask prices of the options at the moment of time $t$ and $sb(t)$ and $sa(t)$ the bid and ask prices of the stock at the moment of time $t$. It is also known that 
Let’s buy an option if the following holds:
$EST(\tau) \ge REAL(0)$.
The trading strategy was developed by Golubnichiy and his colleagues and the results of our models are listed in the following section.
It is with this trading strategy we generated the previously mentioned 57.14\% precision (profitable rate) for CNN Approach 1.1. For the following sections, we will merge our CNN approaches (1.0 and 1.1) into one combined result --- CNN Approach: 51.49\% accuracy, 57.14\% precision, 2.78\% recall.

\section{Result}

In comparison to the pure mathematics approach, QRM, the modified CNN approach yields higher accuracy, and precision, while sacrificing recall. Table 4 lists results from all 3 models developed by Klibanov et.al\cite{SH}, and CNN.

\subsection{Table 4. Final Results on Test Data}
\bigskip
\begin{center}
\begin{tabular}{ |p{4cm}||p{1.75cm}|p{1.75cm}|p{1.75cm}|}
\hline
Method & Accuracy & Precision & Recall\\
\hline
QRM  & 49.77\% & 55.77\% & 52.43\% \\
\hline
Binary Classification & 56.36\% & 59.56\% & 70.22\%  \\
\hline
Regression NN & 55.42\% & 60.32\% & 61.29\% \\
\hline
CNN Approach & 51.49\%  & 57.14\% & 2.78\% \\
\hline
\end{tabular}
\end{center}
\bigskip

These different models all produce various results from the same input data set. However, one cannot simply conclude that one approach is superior to the others, because of the many hidden constraints and limitations of the choice of the data set. features, hyperparameters, etc.   
These results suggest that the CNN model is able to extract relevant information from the input and learn to make predictions. They also emphasize the high volatility of the stock market and the difficulty in simulating the real-world economy with mathematical models. The input stock options are trending toward growth while the outputs are somewhat concentrated. More work can be done to further optimize the model. Since the accuracy is improved compared to QRM, our model might be a better fit for investing in options with both trends (including purchasing options predicted to decrease in value). One possible explanation for the lower recall value is that the stock options data set mostly outputs increasing value, which tunes the predicted trend, resulting in a lowering recall rate. The research team's hypothesis is that the current model may fit better on a different stock options data set. The CNN approach appears promising, and producing better prediction results in future models seems highly possible. Stock option traders may refer to the trading strategy mentioned above and model prediction results to achieve better investment profits.

\section{Future Developments And Applications}
There lies great potential for both developments and applications of the research project. In this section, we introduce some limitations for future improvements.

For the mathematical conceptual approach, more factors such as transaction costs and the magnitude of each stock option’s change could be adopted instead of relying solely on the direction. How to mathematically compute the hedging of various combinations of portfolios is also a challenge. For Machine Learning modeling, different ML models can be developed by changing the algorithms, weights, and layers for current models. We believe that there might exist some negative factors in the data which affect the prediction results. Covid - 19 and the ongoing warfare might increase the instability of the stock market. As for the current CNN model, additional testing and evaluation can be done on different markets, not limited to the selected U.S. companies from a short time range. Lastly, for the trading strategy, risks may be introduced and special circumstances that rock the market could be taken into consideration.

A new project is being developed to improve the program modeling and trading strategies. The project applies Recurrent Neural Network and Long Short-Term Memory (LSTM) to make an additional prediction and a Binomial Asset Pricing / Trading Model to discretely optimize the results. While treating each state as an individual date, we thus justify the model as a stochastic process to reduce the run-time computation from $2^{n}$ to n, whereas n stands for the nth day since the start. We are optimistic that this option forecasting and trading research 2.0 can help resolve partial, if not all, of the up-mentioned constraints.

\section{Summary}
In conclusion, a new machine learning approach is modeled to help predict stock option trends with solutions generated from the Quasi-Reversibility Method and Black-Scholes equation. After processing the Convolutional Neural Networks Machine Learning results by the trading strategy, we concluded a 57.14\% profitable rate for options with an accuracy of 51.49\% and a recall value of 2.78\% for the current CNN approach. Since the accuracy is improved compared to QRM, this model might be a better fit for investmentd options with both trends (including purchasing options predicted to decrease.) Based on the final result of CNN Approach 1, a hypothesis on modeling CNN Approach is established for future improvements. Different parameters and machine learning models will be applied and evaluated to produce new predictions. As introduced in the future developments and applications section, we believe there lies great potential for the current option forecasting model, applying both Mathematical equations and machine learning algorithms.

\subsection{Table 5. Percentages of options with profits/ losses for different methods}
\bigskip
\begin{center}
\begin{tabular}{ |p{3.8cm}||p{3.8cm}|p{3.8cm}|}
\hline
Method & Profitable Options & Options with Loss\\
\hline
QRM  & 55.77\% & 44.23\% \\
\hline
Binary Classification & 59.56\% & 40.44\% \\
\hline
Regression NN & 60.32\% & 39.68\% \\
\hline
CNN Approach & 57.14\%  & 42.86\% \\
\hline
\end{tabular}
\end{center}
\bigskip

Table 5 above summarizes all previous modeling results produced by Dr. Klibanov et.al\cite{SH} and what is presented in this paper. While profitable option rates vary depending on model selections and procedures, the specific choice of time and range of the source data may also have impacts on the final results.

Most significantly, all machine learning models are proven to help improve the stock option forecasting results, which may help investors make better, more profitable, trading decisions.

\end{document}